\pdfoutput=1

\documentclass[11pt,twoside,a4paper,cmspaper,final,collab]{cms-tdr}
\def\svnVersion{264845}\def\cmsCernNo{2013-037}\def\cmsCernDate{\today}\def\cmsMessage{Published in the Journal of High Energy Physics as \href{http://dx.doi.org/10.1007/JHEP10(2014)087}{\doi{10.1007/JHEP10(2014)087.}}}

\begin{document}\cmsNoteHeader{SMP-12-022}

\hyphenation{had-ron-i-za-tion}
\hyphenation{cal-or-i-me-ter}
\hyphenation{de-vices}

\RCS$Revision: 261478 $
\RCS$HeadURL: svn+ssh://svn.cern.ch/reps/tdr2/papers/SMP-12-022/trunk/SMP-12-022.tex $
\RCS$Id: SMP-12-022.tex 261478 2014-09-23 11:17:00Z alverson $

\providecommand{\alphas}{\ensuremath{\alpha_\mathrm{s}}\xspace}
\newcommand{\ytwothree}{\ensuremath{Y_{23}}\xspace}
\newcommand{\njet}{\ensuremath{N_\text{jet}}\xspace}
\newcommand{\taur}{\ensuremath{\tau_{\perp}}\xspace}
\newcommand{\bt}{\ensuremath{B_\text{tot}}\xspace}
\newcommand{\rhoh}{\ensuremath{\rho_\mathrm{H}}\xspace}
\newcommand{\rhos}{\ensuremath{\rho_{\text{tot}}}\xspace}
\newcommand{\trhoh}{\ensuremath{\rho^\mathrm{T}_\mathrm{H}}\xspace}
\newcommand{\trhos}{\ensuremath{\rho^\mathrm{T}_\text{tot}}\xspace}
\newcommand{\nt}{\ensuremath{\hat{n}_\mathrm{T}}\xspace}
\newcommand{\pti}{\ensuremath{p_{\mathrm{T},i}}\xspace}
\newcommand{\ptj}{\ensuremath{p_{\mathrm{T},j}}\xspace}
\newlength\cmsFigWidth\setlength\cmsFigWidth{0.90\columnwidth}
\cmsNoteHeader{SMP-12-022}
\title{Study of hadronic event-shape variables  in multijet final states in pp collisions at $\sqrt{s} = 7\TeV$}

\author[sinp]    {S. Banerjee}
\author[sinp]    {D. Roy}
\author[tifr]    {M. Guchait}
\author[tifr]    {G. Majumder}
\author[sant]    {S. Bhowmik}
\author[sant]    {M. Maity}

\date{\today}

\abstract{
Event-shape variables, which are sensitive to perturbative and
nonperturbative aspects of quantum chromodynamic (QCD) interactions,
are studied in multijet events recorded in proton-proton collisions at
$\sqrt{s} = 7\TeV$. Events are selected with at least one jet with transverse
momentum $\pt>110\GeV$ and pseudorapidity
$\abs{\eta}<2.4$, in a data sample corresponding to integrated luminosities
of up to 5\fbinv. The distributions of
five event-shape variables
in various leading jet \pt ranges are compared to predictions
from different QCD Monte Carlo event generators.
}

\hypersetup{%
pdfauthor={CMS Collaboration},%
pdftitle={Study of hadronic event-shape variables in multijet final states in pp collisions at sqrt(s) = 7 TeV},%
pdfsubject={CMS},%
pdfkeywords={CMS, physics, multijets, event shape}}

\maketitle

\section{Introduction}\label{sec:intro}

Event-shape variables measure the properties of the energy flow in the final states of high energy particle
collisions. Their simplicity combined with their sensitivity to many important signatures of quantum
chromodynamics (QCD)~\cite{jhep0408062,hepph10014082} make them interesting observables
extensively studied in hadronic final states of electron-positron ($\Pep\Pem$) and deep inelastic scattering
(DIS)
collisions~\cite{shapereview,shapeatep,shapeatee}. Such event-shape variables are
theoretically defined in an infrared-  and collinear-safe manner
and can be computed using perturbative techniques. Their measurements have
improved our understanding of many perturbative and nonperturbative
aspects of QCD including the determination of the strong coupling constant
\alphas, details
of parton radiation and hadronization, tests of the colour structure of the
theory, as well as modelling and validation of Monte Carlo (MC) event generators.

Measurements of event-shape variables in hadron-hadron collisions are more complicated than in $\Pep\Pem$ or DIS
collisions, because a larger fraction of the final-state activity is emitted at very forward  pseudorapidities
not covered by the detectors,
and also because the elementary (parton-parton) kinematics cannot be determined as precisely.
These difficulties have led to the redefinition of event-shape variables in the transverse plane,
where the energy flow can be measured with small systematic uncertainty.
A large set of event-shape variables in proton-proton (pp) collisions, which are
sensitive to different aspects of the rich dynamics of the strong interaction
from soft (hadronization) to hard (multijet radiation) scales has been
proposed~\cite{jhep0408062,hepph10014082}.
These variables are normalized to the sum of the measured transverse
momenta (\pt) of all reconstructed objects in the event to reduce the
systematic uncertainty due to the jet energy scale.

Previous studies of event-shape variables at hadron colliders include those of the CDF experiment at
the Tevatron~\cite{evtshape_cdf},
and early measurements at the LHC~\cite{evtshape_cms1,evtshape_atlas2}.
More recently event-shape variables have been studied in the associated production of Z bosons with jets~\cite{evtshape_cms2}.
In the previous analysis of the CMS experiment with 3.2\pbinv of data~\cite{evtshape_cms1},
the transverse thrust and thrust minor variables were studied
to improve the modelling
of multijet production in MC generators. This study is expanded here using
a larger data set corresponding to an integrated luminosity of 5\fbinv
in pp collisions at $\sqrt{s} = 7\TeV$ with
an expanded set of five event-shape variables~\cite{jhep0408062,hepph10014082}:
the transverse thrust, jet broadening, jet mass (both total and in the transverse plane), and the third-jet
resolution parameter. The significant increase in luminosity allows the measurement of variables with
three jets, not accessible with earlier data, and the latter four
observables are analysed in CMS for the first time. Therefore this
analysis is sensitive to features of the event generators that were
not probed in the previous CMS result.

The paper is organized as follows. In Section~\ref{sec:cmsdet},
elements of the CMS detector relevant to this analysis are described.
Section~\ref{sec:def} introduces the event-shape variables studied in this work.
The data and MC simulated event samples are summarized in
Section~\ref{sec:evtsel} along with the event selection criteria.
Section~\ref{sec:system} describes the unfolding
technique employed and the propagated systematic uncertainties.
Section~\ref{sec:results} compares the five event-shape
distributions in data with several QCD event generators.
The results are summarized in Section~\ref{sec:summ}.

\section{The CMS detector}\label{sec:cmsdet}

The CMS experiment~\cite{cmsdet} uses a right-handed coordinate system, with
the origin at the nominal interaction point, the $x$ axis pointing to the
centre of the LHC ring, the $y$ axis pointing vertically up (perpendicular to the plane of the LHC ring),
and the $z$ axis along the anticlockwise-beam direction.
The polar angle $\theta$ is measured from the positive $z$ axis and the
azimuthal angle $\phi$ is measured in the $x$-$y$ plane in radians. Pseudorapidity is
defined as $\eta = - \ln [\tan (\theta/2)] $.

The central feature of the CMS detector is a superconducting solenoid of 6\unit{m}
internal diameter, providing a magnetic field of 3.8\unit{T}. Within the field volume,
there are silicon pixel and strip trackers, a lead tungstate crystal
electromagnetic calorimeter (ECAL), and a sampling hadron calorimeter
made up of layers of brass plates and plastic scintillators. The calorimeters
provide coverage in pseudorapidity up to $\abs{\eta} = 3.0$.
A preshower detector consisting of two planes of silicon sensors interleaved with lead is
located in front of the ECAL at $1.7 <\abs{\eta}< 2.6$. An iron and quartz fiber
Cherenkov hadron calorimeter covers pseudorapidities $3.0 <\abs{\eta} < 5.0$.
The muons are measured in the pseudorapidity range $\abs{\eta} < 2.4$, with detection planes
made using three technologies: drift tubes, cathode strip chambers, and resistive plate chambers.

The particle-flow (PF)
algorithm~\cite{PFJet1,PFJet2} combines information on charged
particles from the tracking system, energy deposits in the electromagnetic
and hadron calorimeters, as well as signals in the preshower detector and
muon systems to assign a four-momentum vector to particles, \ie
$\gamma$, \Pe$^\pm$, \Pgm$^\pm$, charged, and neutral hadrons.
Jets are reconstructed using these particles. The energy calibration of individual
particle types is performed separately. At the PF level, the jet constituents are almost fully calibrated and
require only a small correction (less than 10\%)~\cite{jetenergyreso} due to tracking inefficiencies and
threshold effects. The jet clustering is performed using the anti-\kt clustering
algorithm~\cite{antikt,fastjet} with a distance parameter $R=0.5$. The jets are ordered by descending
\pt with $p_{\mathrm {T,1}}$ and $p_{\mathrm {T,2}}$ representing the transverse
momenta of the leading and the second leading jets, respectively.

\section{Event-shape variables}\label{sec:def}
Five event-shape variables are analysed in this paper:
the transverse thrust $\taur$, the total
jet broadening $\bt$, the total jet mass $\rhos$,
the total transverse jet mass $\trhos$
and the third-jet resolution parameter $\ytwothree$.
In the formulae below, $\pti$, $\eta_i$, and $\phi_i$ represent the
transverse momentum, pseudorapidity, and azimuthal angle of the $i$th jet,
and $\nt$ is the unit vector that maximizes the sum of the projections of
$\vec{p}_{{\mathrm T},i}$.
The transverse thrust axis $\nt$ and the beam form the so-called event plane.
Based on the direction of $\nt$,
the transverse region is separated into an upper side ${\mathcal{C}}_\mathrm{U}$, consisting
of all jets with $\vec{p}_\mathrm {T}\cdot\nt> 0$, and a lower side
${\mathcal{C}}_\mathrm {L}$, with $\vec{p}_\mathrm {T}\cdot\nt< 0$. The
jet broadening and third-jet resolution variables require at least three selected jets,
whereas the calculation of other variables requires at least two jets.
 The $\nt$ vector is defined only up to a global sign -
choosing one sign or the other has no consequence since it
simply exchanges the upper and lower event regions.

\begin{description}

\item[Transverse thrust:]
The event thrust observable in the transverse plane is defined as
\begin{equation}
 \taur  \equiv  1 - \max_{\nt} \frac{\sum_i \abs{\vec{p}_{\mathrm {T},i} \cdot \nt }}
                                    {\sum_i \pti}.
\end{equation}
This variable probes the hadronisation process
and is sensitive to the modelling of two-jet and multijet
topologies.
In this paper ``multijet" refers to ``more-than-two-jet".
In the limit of a perfectly balanced two-jet event,
$\taur$ is zero,
while in isotropic multijet events it amounts to $(1-2/\pi)$.

\item[Jet broadening:] The
pseudorapidities and the azimuthal angles of the axes
for the upper and lower event regions are defined by
\begin{align}
 \eta_X & \equiv  \frac{\sum_{i\in{\mathcal{C}}_X} \pti\,\eta_i}
                        {\sum_{i\in{\mathcal{C}}_X} \pti}, \\
 \phi_X & \equiv  \frac{\sum_{i\in{\mathcal{C}}_X} \pti\,\phi_i}
                        {\sum_{i\in{\mathcal{C}}_X} \pti},
\end{align}
where $X$ refers to upper (U) or lower (L) side. From these, the jet broadening variable in each region is defined as
\begin{equation}
 B_{X}  \equiv  \frac{1}{2\,P_{\,\mathrm T}} \sum_{i\in{\mathcal{C}}_X}
\pti
                \sqrt{(\eta_i -\eta_X)^2 + (\phi_i - \phi_X)^2},
\end{equation}
where $P_{\,\mathrm T}$ is the scalar sum of the transverse momenta of all the
jets. The total jet broadening is then defined as
\begin{equation}
 \bt  \equiv  B_{\mathrm U} + B_{\mathrm L}.
\end{equation}

\item[Jet masses:]
The normalized squared invariant mass of the jets in the upper and lower regions of the event is
defined by
\begin{equation}
\rho_X  \equiv  \frac{M_{X}^2}{P^2},
\end{equation}
where $M_{X}$ is the invariant mass of the constituents of the jets in the region $X$,
and $P$ is the scalar sum of the momenta of all constituents in both sides.

The jet mass variable is defined as the sum of the masses in the upper and lower regions,
\begin{equation}
\rhos \equiv \rho_{\mathrm U} + \rho_{\mathrm L}.
\end{equation}
The corresponding jet mass in the transverse plane,
$\trhos$, is also similarly calculated in transverse plane.

\item[Third-jet resolution parameter:]
The third-jet resolution parameter is defined as
\begin{equation}
\ytwothree \equiv \frac{{\min} (p_{\mathrm {T,3}}^2\,,\:
[\min(\pti\,,\, \ptj)^2 \times (\Delta R_{ij})^2/ R^2])} {P_{12}^2},
\end{equation}
where $i,j$ run over all three jets,
$(\Delta R_{ij})^2 = (\eta_i - \eta_j)^2 + (\phi_i - \phi_j)^2$, and
$p_{\mathrm {T},3}$ is the transverse momentum of the third jet in the event.
If there are more than three jets in the event, they are iteratively merged
using the \kt algorithm~\cite{kt4a,kt4b} with a distance parameter $R=0.6$.
To compute $P_{12}$, three jets are merged into two using the procedure
described above and $P_{12}$ is then defined as the scalar sum of the transverse
momenta of the two remaining jets.

The $\ytwothree$ variable estimates the relative strength of the \pt of the third jet with respect to the other two jets.
It vanishes for two-jet events, and a nonzero value of $\ytwothree$ indicates the presence of hard
parton emission, which tests the parton showering model of QCD event generators.
A test like this is
less sensitive to the details of the underlying event (UE) and parton hadronization models
than the other event-shape variables~\cite{hepph10014082}.

\end{description}

\section{Event selection and Monte Carlo samples}\label{sec:evtsel}

This analysis extends the phase space compared to the previous study~\cite{evtshape_cms1}
to $\abs{\eta}< 2.4$, and considers several different \pt ranges for the leading jet.
The events used are collected with single-jet
triggers, which are reconstructed from calorimeter information only,
where the \pt of at least one jet is above a certain threshold, $p_{\mathrm{T,th}}$.
Events are divided into five bins of $p_{\mathrm {T,1}}$ where each bin uses data from
one trigger path.
The choice of $p_{\mathrm {T,1}}$ ranges (Table~\ref{tab:dataset}) has been
determined by the trigger criteria, while the \pt threshold (${>}30\GeV$) for
the other jets and their geometric acceptance ($\abs{\eta}< 2.4$) are restricted to
give the good jet energy scale and resolution.
Spurious jets, which are due to noise in the calorimeters or other noncollision
backgrounds, are eliminated using jet quality criteria, \eg
jets must consist of at least two particles, including at least one charged
hadron, and not more than 99\% of the jet energy may be carried
by neutral hadrons alone, or by photons alone. % and are removed.
These jets that do not satisfy the identification requirements are not
included in the calculation of the event-shape variables.

An event is discarded if
\begin{itemize}
\item any one of the two highest-\pt jets in the event lies outside the central region ($\abs{\eta} < 2.4$);
 for the measurement of $\bt$ and $\ytwothree$ a third jet satisfying the jet selection criteria is required within
the same detector acceptance region;
\item any one of the two highest-\pt jets is spurious;
\item all selected jets of an event lie only on one side of the line perpendicular to $\nt$.
This criterion ensures that events will be rejected if jets are missed in the forward direction.
Events of interest for this analysis should be well-balanced in \pt and hence have jets on both sides of
this line.
\end{itemize}

Table~\ref{tab:dataset} shows the numbers of events, as well as the
fractions of events with two, three, four, or more jets, for various ranges of the leading jet
$p_{\mathrm {T,1}}$, along with the effective integrated luminosity for each data sample.
The effective luminosities differ due to variations of the prescale factor of the trigger paths associated with each $p_{\mathrm {T,1}}$.
The average number of additional pp interactions per bunch crossing (pileup) on the collected dataset is
$\approx$8. The effect of pileup in the distributions of
event-shape variables has been studied by
grouping the events in different ranges of number of reconstructed primary
vertices, and no bias has been found. This is expected due to fact that after the
jet energy calibration, there is no residual pileup dependence.

\begin{table}[htbp!]
\topcaption{Characteristics of the data samples selected for this analysis, in categories of leading jet transverse momentum $p_{\mathrm {T,1}}$: effective
                integrated luminosity, selected number of events, and relative abundances of the numbers of selected jets, $\njet$, for jets
		with $\pt > 30\GeV$ and $\abs{\eta} < 2.4$.}
\begin{center}\begin{tabular}{cccrcccc} \hline
 Trigger & Range of  &  Luminosity      & Number   & \multicolumn{4}{c}{Fraction of events (\%)} \\  \cline{5-8}
$p_{\mathrm{T,th}}$ (\GeVns) & $p_{\mathrm{T,1}}$ (\GeVns)           &   &  of events &  $\njet=2$    & $\njet=3$ & $\njet=4$ & $\njet {>} 4$\\ \hline
60 & 110--170          &    0.403\pbinv     &  96833  &  57.9 & 32.6 & 7.8  & 1.7  \\
110 & 170--250         &    7.15\pbinv	  &  228854 &  43.0 & 37.7 & 14.6 & 4.7 \\
190 & 250--320         &    153\pbinv	  &  601554 &  34.8 & 37.8 & 19.1 & 8.3 \\
240 & 320--390         &    521\pbinv      &  497827 &  31.0 & 37.0 & 21.2 & 10.9\\
300 & $>$390          &    4.98\fbinv     &  2234304&  28.4 & 35.6 & 22.5 & 13.5\\
\hline
\end{tabular}\end{center}
\label{tab:dataset}
\end{table}

Four MC generators, \PYTHIA 6.426 (\PYTHIA6)
~\cite{pythia6}, \PYTHIA 8.153 (\PYTHIA8)~\cite{pythia8},
\HERWIGpp 2.5.0 (\HERWIGpp)~\cite{herwig++}, and
\MADGRAPH 5.1.5.7 (\MADGRAPH)~\cite{madgraph}
are chosen to generate multijet events.
Particles with a lifetime larger than 30\unit{ps} are declared stable and handled by the full
CMS detector simulation based on \GEANTfour~\cite{geant4}.
These generators reproduce  the single differential jet spectra
measured at the LHC~\cite{CMS:2011ab,Chatrchyan:2012gwa,Chatrchyan:2013qza}.
The simulated events are then reconstructed in the same way
for real data. The MC simulations are also used to obtain the unfolding
corrections, described in the next section, and to estimate the
associated uncertainties.

Events are generated with \PYTHIA6 using three different models:
(i) D6T~\cite{d6t}, which uses virtuality-ordered parton showering (PS)
and is based on Tevatron data; and two models that use \pt-ordered PS:
(ii) Perugia-P0~\cite{perugia} based on LEP and Tevatron data,
and (iii) Z2~\cite{z2tune} based on CMS data collected at $\sqrt{s} = 900\GeV$ and $7\TeV$.
The generator \PYTHIA8 uses \pt-ordered PS, an UE description
based on the multiple parton interaction (MPI) model of \PYTHIA6 interleaved with initial and
final state radiation, and the
tune4C~\cite{pythia8tune2c} settings.
The \HERWIGpp generator is run with tune23 settings,
where the PS evolution is based on angular ordering and an eikonal MPI
model for the UE.
Finally, the \MADGRAPH MC employs matrix element (ME)
calculations to generate events with two to four partons
plus \PYTHIA6-tuneZ2 for the PS and UE. The MLM matching procedure~\cite{mlm} is imposed to avoid a
double counting of jets between the ME and PS, for a minimum jet \pt threshold of 20\GeV.

All MC generators use the CTEQ6L1 parametrization as the choice of parton distribution function (PDF), except for
Perugia-P0 which uses CTEQ5L~\cite{cteq61}.

\section{Unfolding and systematic uncertainties}\label{sec:system}

Jets at generator level are defined as a collection of stable particles
with the same kinematic criteria used for the real data.
The distribution of a
variable obtained using parton- and detector-level
information differs because of the finite energy and angular
resolutions of the experimental apparatus.
In order to correct the measured distributions for bin migrations
due to detector effects,
a response matrix is constructed with simulated events.
The D'Agostini method~\cite{bayes} is employed to unfold the experimental
data, using the response matrix obtained from \PYTHIA6-tuneZ2, \PYTHIA8, and \MADGRAPH samples.
Although the results are consistent for the generators, small differences ($<$3\%) are observed, which are taken
as a systematic uncertainty.
Another source of systematic uncertainty in the unfolding procedure is the choice of the unsmearing method.
A regularized unfolding method based on singular value decomposition (SVD) of the response matrix~\cite{SVD}
is also used as a consistency check.
The difference between the D'Agostini and SVD unfolding methods
is less than 5\% for the $\taur$ distribution.
It can be as high as 20\% for the distributions of other event-shape variables, which require
more than three jets in the event, mainly as a consequence of the lower number of
events in the lower ranges of $p_{\mathrm{T},1}$.

Other sources of systematic uncertainty include the finite jet energy and
angular resolutions and the jet energy scale~\cite{jetenergyreso}.
In order to propagate the uncertainties due to the jet resolutions, the unfolded
response matrix is obtained with jets randomly spread at the generator level with increasing and decreasing values
of the resolution parameters. The corresponding differences in the unfolded data distributions
are considered as systematic uncertainties, which are found to be less than 2\% in most
cases, but can be as high as 5\% in some corners of phase space.
Similarly, the jet energy scale is increased and decreased by one standard deviation with respect to its
central value and the unfolded distributions are compared with the nominal one to estimate the effect of this
scale correction, the resulting uncertainty is less than 3\%.
The effect of pileup on the event-shape variables is found to be negligible.

\section{Results}\label{sec:results}

\begin{figure}[htbp]
\begin{center}
\includegraphics[width=\cmsFigWidth]{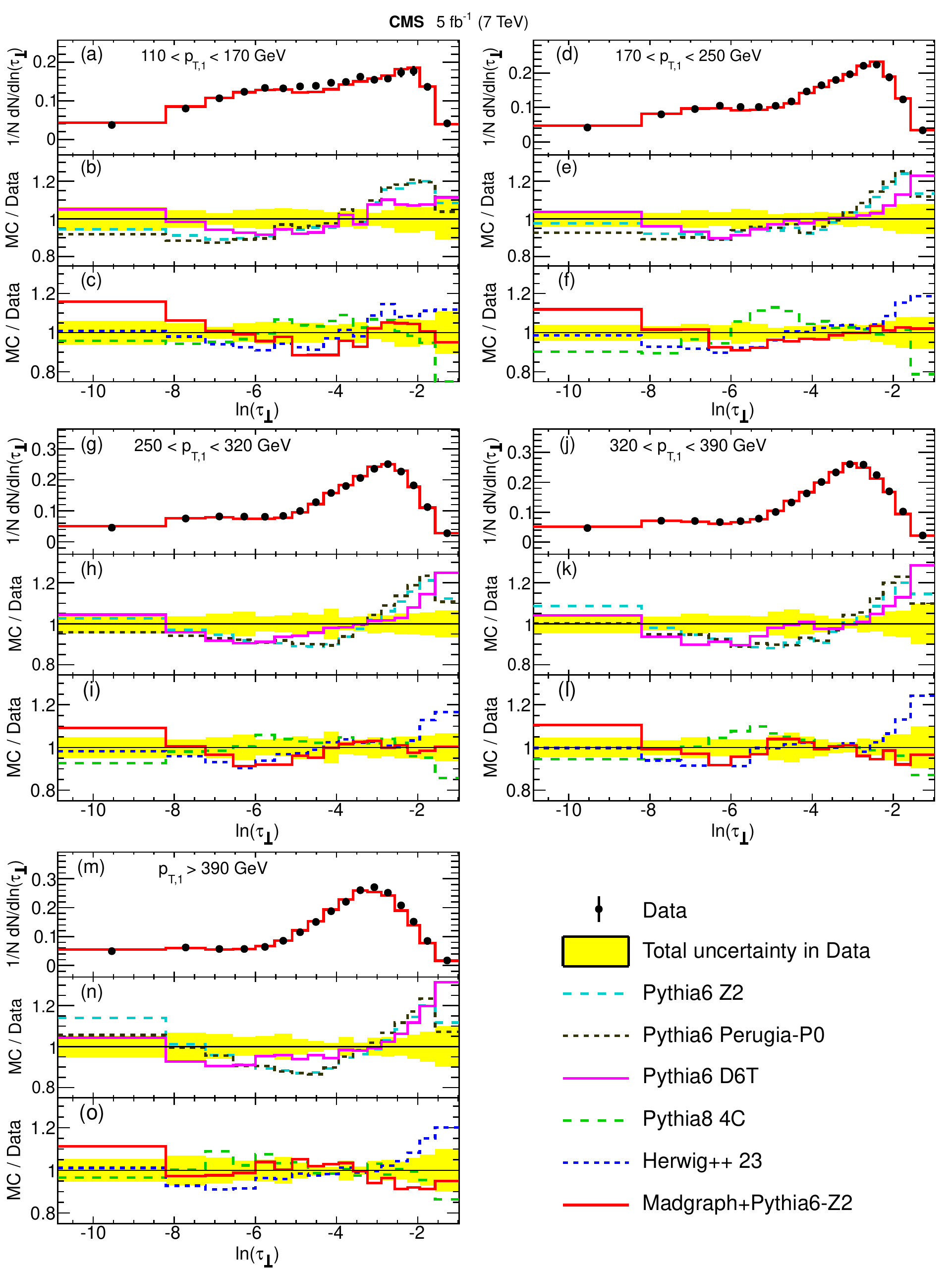}
\end{center}
\caption{(a,d,g,j,m) Comparison between the transverse thrust $\taur$
distributions in data and  \MADGRAPH{}+\PYTHIA6-Z2 event generator in five different ranges
of $p_{\mathrm {T,1}}$. The error bars around the
data points indicate the statistical uncertainties in data.
         The panels (b,e,h,k,n) show the ratios of different models of the \PYTHIA6 event
         generator over data
         in each momentum range and panels (c,f,i,l,o) show the ratios for other generators.
         The shaded bands represent statistical and systematic uncertainties in data.}
\label{fig:finalhlt_thrust}
\end{figure}

\begin{figure}[htbp]
\begin{center}
\includegraphics[width=\cmsFigWidth]{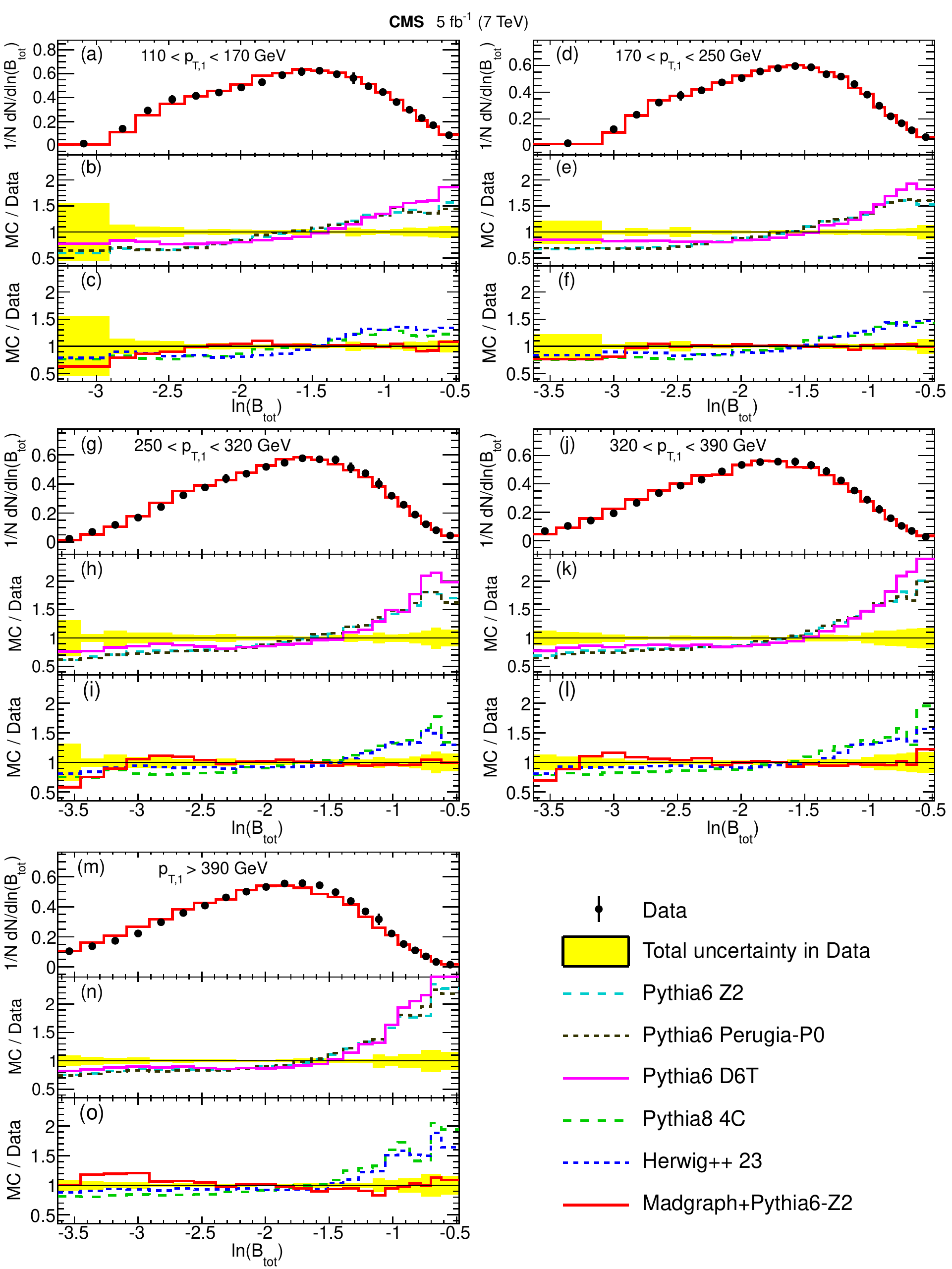}
\end{center}
\caption{Comparison between the jet broadening $\bt$ distributions in data and various Monte Carlo models.
The \pt bins and other details are the same as in Fig.~\ref{fig:finalhlt_thrust}.}
\label{fig:finalhlt_jetbroadening}
\end{figure}

\begin{figure}[htbp]
\begin{center}
\includegraphics[width=\cmsFigWidth]{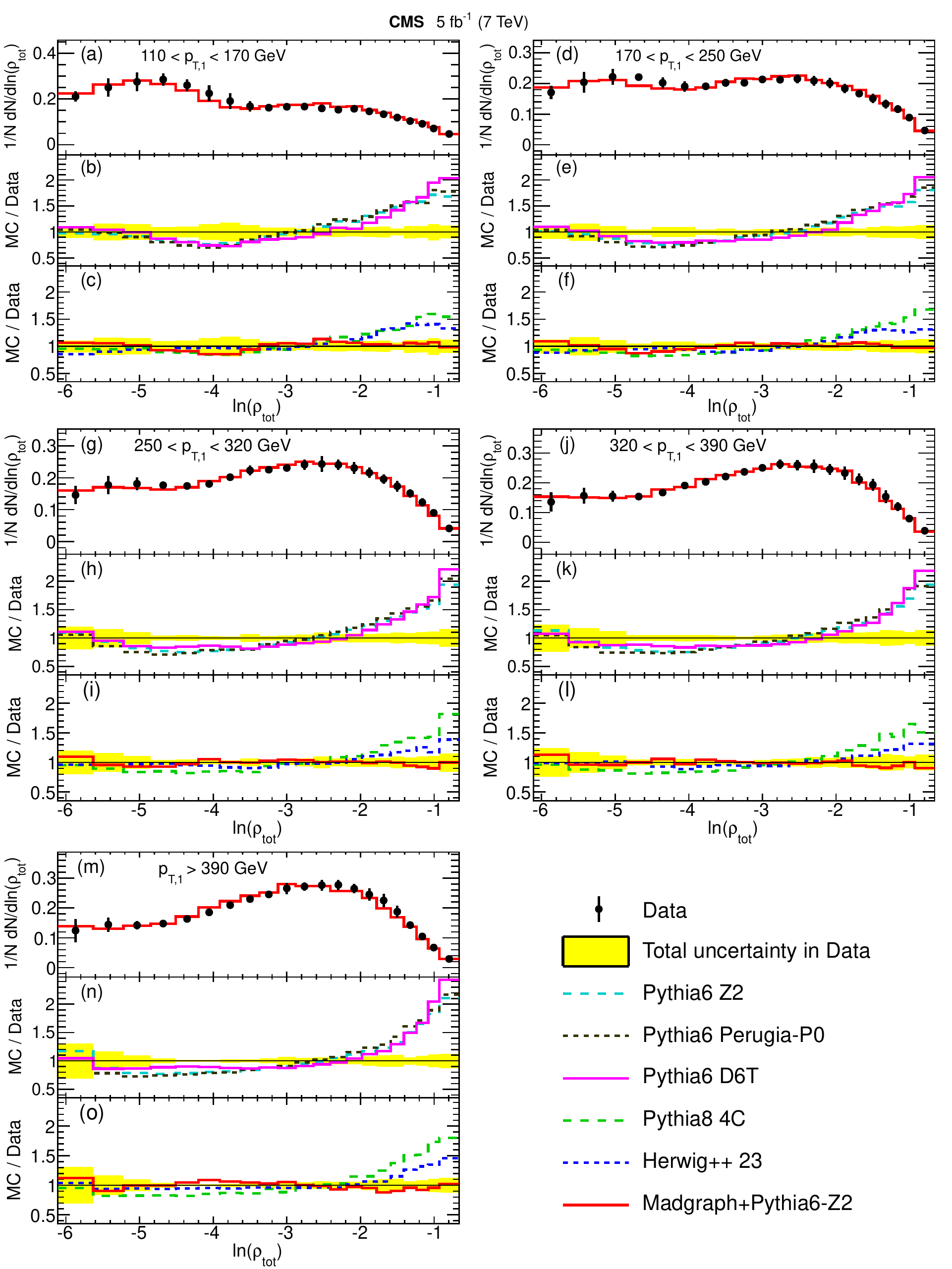}
\end{center}
\caption{Comparison between the total jet mass $\rhos$ distributions in data and various Monte Carlo
  models. The \pt bins and other details are the same as in Fig.~\ref{fig:finalhlt_thrust}.}
\label{fig:finalhlt_jetmass}
\end{figure}

\begin{figure}[htbp]
\begin{center}
\includegraphics[width=\cmsFigWidth]{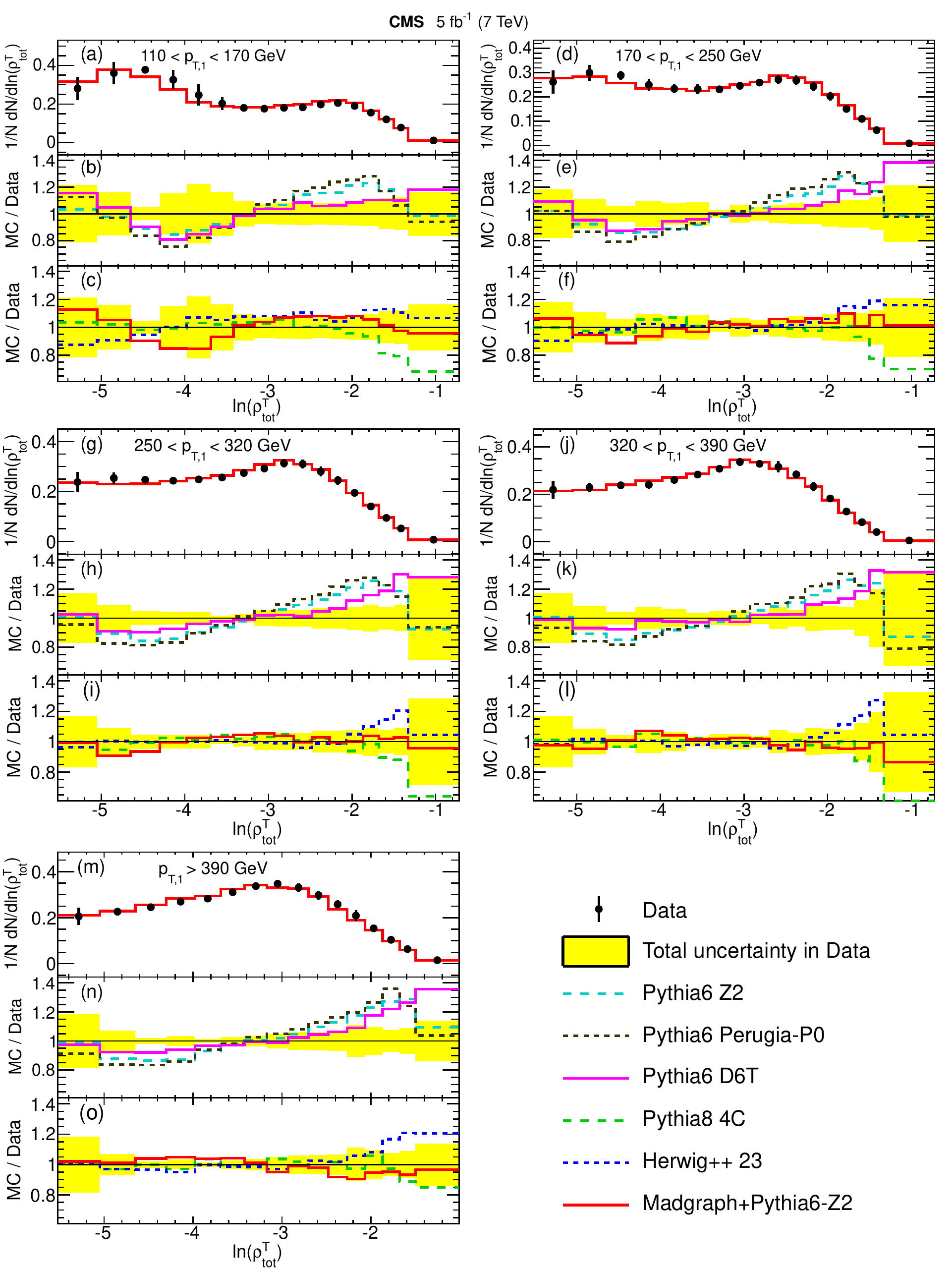}
\end{center}
\caption{Comparison between the total jet transverse mass $\trhos$ distributions in data and various Monte
  Carlo models. The \pt bins and other details are the same as in Fig.~\ref{fig:finalhlt_thrust}.}
\label{fig:finalhlt_jettransvmass}
\end{figure}

\begin{figure}[htbp]
\begin{center}
\includegraphics[width=\cmsFigWidth]{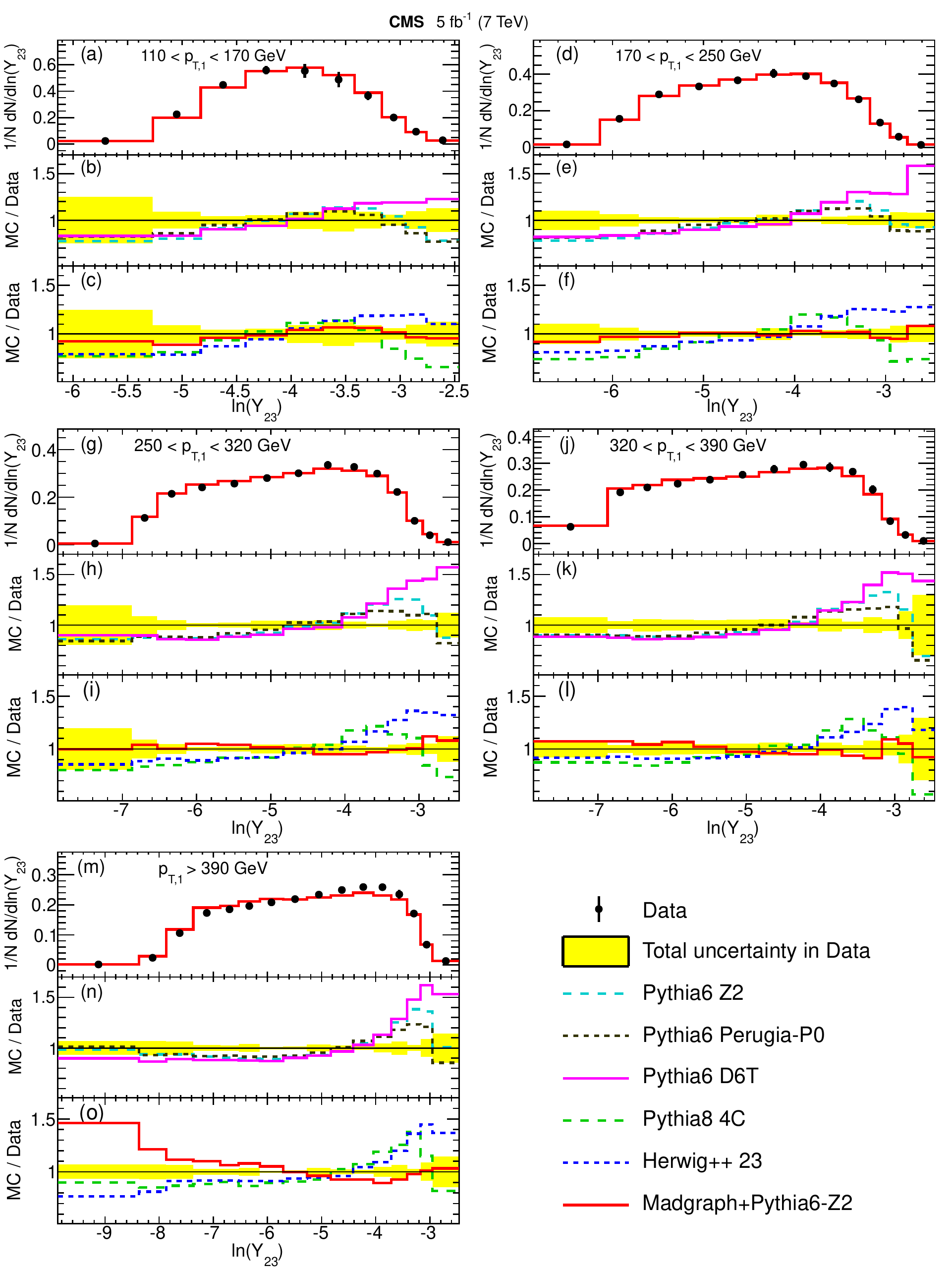}
\end{center}
\caption{Comparison between the third-jet resolution parameter $\ytwothree$ in data and various Monte Carlo
  models. The \pt bins and other details are the same as in Fig.~\ref{fig:finalhlt_thrust}.}
\label{fig:finalhlt_jety23}
\end{figure}

The distributions of the logarithms of the five event-shape variables analysed
 ($\taur$, $\bt$, jet mass (both $\rhos$ and $\trhos$), and $\ytwothree$) are shown in
Figs.~\ref{fig:finalhlt_thrust}--\ref{fig:finalhlt_jety23} for the five leading jet \pt ranges listed in
Table~\ref{tab:dataset}.
All distributions are unfolded, normalized to unitary; they are compared to
the predictions from the six generator models.
The error bars around the data points indicate the statistical uncertainties and the shaded bands
represent the sum in quadrature of statistical and systematic  uncertainties. The corresponding ratios
between the model predictions and the data are shown in the lower plots.
The distributions are plotted in a logarithmic horizontal axis scale so that the details at small values of the
event-shape variables are also visible.

Overall, the models tend to reproduce
 the transverse thrust $\taur$, total transverse jet mass $\trhos$, and third-jet resolution parameter
$\ytwothree$ distributions better  than the total jet mass $\rhos$ and jet broadening $\bt$ ones.
The model that consistently reproduces all the distributions within the uncertainties
is the \MADGRAPH matrix-element calculator combined with \PYTHIA6-tuneZ2 for the PS and UE.

Similar data-MC comparisons are performed using these different jet clustering
algorithms: (i) anti-\kt with a distance parameter $R = 0.7$, (ii) \kt with
a distance parameter $R = 0.4$, and (iii) energy deposits using calorimeter
information only instead of PF candidates.
In all cases, the results are similar and in agreement with
each other. Also, the effect due to the choice of a particular PDF set in the
MC predictions
has been estimated using the MSTW2008lo68cl set~\cite{mstw2008,mstw2009}.
A negligible effect has been found by varying the PDF eigenvalues within one standard deviation.

The transverse thrust variable $\taur$ (Fig.~\ref{fig:finalhlt_thrust}) is insensitive to the longitudinal component of the
particles' momenta, and thus to the modelling of MPI and colour connection between soft scatters and beam
remnants. The data-MC agreement for this observable is at the 5--10\% level for all \pt bins
except at the highest $\taur$ where differences as large as 20\% are observed.
The agreement is better than the
other event-shape variables, which are more sensitive to MPI and colour connection effects.
The $\taur$ distributions also reveal that the predictions for the lower \pt bins from
\PYTHIA8, \HERWIGpp, \MADGRAPH, and \PYTHIA6 (with model~D6T) are
closer to the data than the ones from \PYTHIA6 with model Z2 and Perugia-P0.

The jet broadening distribution (Fig.~\ref{fig:finalhlt_jetbroadening}) is poorly described by
all the models at both low and high $\bt$ values except for the \MADGRAPH generator.
This variable is insensitive to the UE and hadronization details, but a precise modelling of the ME
and PS is crucial in order to correctly predict its distribution.
Both model ingredients are expected to be more adequately described in \MADGRAPH,
where the multijet final-states are directly obtained from the hard ME calculations,
unlike \PYTHIA and \HERWIGpp parton showers, which work best for 2$\to$2 processes.
In addition, the jet broadening is sensitive to colour coherence effects, which have an improved
description~\cite{Chatrchyan:2013fha} in the current version of \HERWIGpp,
which explains the best relative agreement of this model compared to all \PYTHIA models.
Similar arguments are also applicable for the total jet mass $\rhos$ and the third-jet resolution parameter $\ytwothree$.

The total jet mass $\rhos$ distribution (Fig.~\ref{fig:finalhlt_jetmass}) shows a
similar behaviour between the measurement and the different model predictions as observed
for the jet broadening case. \MADGRAPH and \HERWIGpp reproduce this observable better than the
various \PYTHIA models.
This variable is more sensitive to (initial-state) forward radiation than the jet
broadening~\cite{hepph10014082}, which indicates that such QCD emission is adequately described in the former two models.

The transverse jet mass $\trhos$ distributions (Fig.~\ref{fig:finalhlt_jettransvmass}) show agreement between
data and predictions within 20\%, which is better than that seen for the total jet mass $\rhos$.
This is expected for $\rhos$, because transverse variables are less sensitive to the longitudinal energy
flow~\cite{hepph10014082} and to colour connection effects. Among the \PYTHIA6 models,
D6T one shows the best agreement with the data.

The third-jet resolution parameter $\ytwothree$ distribution (Fig.~\ref{fig:finalhlt_jety23}) is
sensitive to the properties of multijet emission and it is robust with respect to the modelling
of the UE and hadronization.
The \MADGRAPH generator shows again, for such a ME-sensitive observable,
a better data-model agreement than the rest of MC simulations.

\section{Summary}\label{sec:summ}

An extended set of five event-shape variables (the transverse thrust $\taur$,
the jet broadening $\bt$,  the total jet mass $\rhos$, the total transverse jet mass $\trhos$,
and the third-jet resolution parameter $\ytwothree$)
have been studied in multijet final states measured in pp collisions at $\sqrt{s}=7\TeV$.
Such observables are sensitive to perturbative and nonperturbative aspects of QCD,
and allow the validation of hadronic event generators.
The experimental distributions have been measured in five different ranges of leading jet
transverse momenta from $110 <\pt < 170\GeV$ up to $\pt>390\GeV$,
and compared to the predictions of six different event generators.

For the transverse thrust, all generators show an overall agreement with the data within 10\%, with
\PYTHIA8 and \HERWIGpp exhibiting a better agreement than the others.
A 20\% level of agreement is also found for the total transverse jet mass distributions.
However, event-shape variables that are more sensitive to the longitudinal energy flow (such as
the total jet mass) or to hard parton emissions (such as the jet broadening) show a larger discrepancy
between data and parton shower MC simulations.
The predictions of \PYTHIA6-D6T show better agreement with
data for the event-shape variables that make use only of the jet
\pt, but have worse agreement for $\ytwothree$ compared to other \PYTHIA6 models.
The modelling of colour connection between the soft scatters and beam remnants, and initial- and final-state
radiations are the major sources of differences between the various QCD event generators.
The generator that consistently reproduces all distributions within the uncertainties is the
\MADGRAPH matrix-element calculator combined with \PYTHIA6-tuneZ2 for multiparton interactions
and parton showering and hadronization. The study of infrared- and collinear-safe event-shape
variables presented here provides detailed information to further improve the
modelling
of parton radiation and hadronization in event generators for high energy hadronic collisions.

\section{Acknowledgments}

We congratulate our colleagues in the CERN accelerator departments for the excellent performance of the LHC and thank the technical and administrative staffs at CERN and at other CMS institutes for their contributions to the success of the CMS effort. In addition, we gratefully acknowledge the computing centres and personnel of the Worldwide LHC Computing Grid for delivering so effectively the computing infrastructure essential to our analyses. Finally, we acknowledge the enduring support for the construction and operation of the LHC and the CMS detector provided by the following funding agencies: BMWFW and FWF (Austria); FNRS and FWO (Belgium); CNPq, CAPES, FAPERJ, and FAPESP (Brazil); MES (Bulgaria); CERN; CAS, MoST, and NSFC (China); COLCIENCIAS (Colombia); MSES and CSF (Croatia); RPF (Cyprus); MoER, ERC IUT and ERDF (Estonia); Academy of Finland, MEC, and HIP (Finland); CEA and CNRS/IN2P3 (France); BMBF, DFG, and HGF (Germany); GSRT (Greece); OTKA and NIH (Hungary); DAE and DST (India); IPM (Iran); SFI (Ireland); INFN (Italy); NRF and WCU (Republic of Korea); LAS (Lithuania); MOE and UM (Malaysia); CINVESTAV, CONACYT, SEP, and UASLP-FAI (Mexico); MBIE (New Zealand); PAEC (Pakistan); MSHE and NSC (Poland); FCT (Portugal); JINR (Dubna); MON, RosAtom, RAS and RFBR (Russia); MESTD (Serbia); SEIDI and CPAN (Spain); Swiss Funding Agencies (Switzerland); MST (Taipei); ThEPCenter, IPST, STAR and NSTDA (Thailand); TUBITAK and TAEK (Turkey); NASU and SFFR (Ukraine); STFC (United Kingdom); DOE and NSF (USA).

Individuals have received support from the Marie-Curie programme and the European Research Council and EPLANET (European Union); the Leventis Foundation; the A. P. Sloan Foundation; the Alexander von Humboldt Foundation; the Belgian Federal Science Policy Office; the Fonds pour la Formation \`a la Recherche dans l'Industrie et dans l'Agriculture (FRIA-Belgium); the Agentschap voor Innovatie door Wetenschap en Technologie (IWT-Belgium); the Ministry of Education, Youth and Sports (MEYS) of the Czech Republic; the Council of Science and Industrial Research, India; the HOMING PLUS programme of Foundation for Polish Science, cofinanced from European Union, Regional Development Fund; the Compagnia di San Paolo (Torino); the Consorzio per la Fisica (Trieste); MIUR project 20108T4XTM (Italy); the Thalis and Aristeia programmes cofinanced by EU-ESF and the Greek NSRF; and the National Priorities Research Program by Qatar National Research Fund.
\bibliography{auto_generated}   % will be created by the tdr script.

\cleardoublepage \appendix\section{The CMS Collaboration \label{app:collab}}\begin{sloppypar}\hyphenpenalty=5000\widowpenalty=500\clubpenalty=5000\textbf{Yerevan Physics Institute,  Yerevan,  Armenia}\\*[0pt]
V.~Khachatryan, A.M.~Sirunyan, A.~Tumasyan
\vskip\cmsinstskip
\textbf{Institut f\"{u}r Hochenergiephysik der OeAW,  Wien,  Austria}\\*[0pt]
W.~Adam, T.~Bergauer, M.~Dragicevic, J.~Er\"{o}, C.~Fabjan\cmsAuthorMark{1}, M.~Friedl, R.~Fr\"{u}hwirth\cmsAuthorMark{1}, V.M.~Ghete, C.~Hartl, N.~H\"{o}rmann, J.~Hrubec, M.~Jeitler\cmsAuthorMark{1}, W.~Kiesenhofer, V.~Kn\"{u}nz, M.~Krammer\cmsAuthorMark{1}, I.~Kr\"{a}tschmer, D.~Liko, I.~Mikulec, D.~Rabady\cmsAuthorMark{2}, B.~Rahbaran, H.~Rohringer, R.~Sch\"{o}fbeck, J.~Strauss, A.~Taurok, W.~Treberer-Treberspurg, W.~Waltenberger, C.-E.~Wulz\cmsAuthorMark{1}
% ...
\vskip\cmsinstskip
% ...
1:~~Also at Vienna University of Technology, Vienna, Austria\\
2:~~Also at CERN, European Organization for Nuclear Research, Geneva, Switzerland\\
% ...

\end{sloppypar}
\end{document}